\ifdefined\WORDCOUNT
\documentclass[superscriptaddress,twocolumn,preprintnumbers,amsmath,amssymb,prl,nofootinbib,groupedaddress]{revtex4-1}
\else
\documentclass[superscriptaddress,twocolumn,showpacs,preprintnumbers,amsmath,amssymb,prl,groupedaddress]{revtex4-1}
\fi
\usepackage[table]{xcolor}

\newcolumntype{s}{>{\columncolor[HTML]{AAACED}} p{3cm}}

\usepackage{amssymb}
\usepackage{epsfig,amsmath,graphics}
\usepackage{epstopdf}
\usepackage{graphicx}
\usepackage{color}
\usepackage{gensymb}

\usepackage[utf8x]{inputenc}
\newcommand{\ket}[1]{\ensuremath{\left|#1\right>}}

\usepackage{natbib}
\usepackage{graphicx}

\begin{document}

\title{Atom-interferometric test of the equivalence principle at the $10^{-12}$ level}

\author{Peter Asenbaum}
\thanks{These authors contributed equally to this work.}
\author{Chris Overstreet}
\thanks{These authors contributed equally to this work.}
\author{Minjeong Kim}
\author{Joseph Curti}
\author{Mark A. Kasevich}
\email{kasevich@stanford.edu}
\affiliation{Department of Physics, Stanford University, Stanford, California 94305}
\date{June 25, 2020}

\begin{abstract}
Does gravity influence local measurements?
We use a dual-species atom interferometer with $2\,\text{s}$ of free-fall time to measure the relative acceleration between $^{85}$Rb and $^{87}$Rb wave packets in the Earth's gravitational field.  Systematic errors arising from kinematic differences between the isotopes are suppressed by calibrating the angles and frequencies of the interferometry beams.  We find an E\"otv\"os parameter of $\eta = [1.6\; \pm\; 1.8\; \text{(stat)}\; \pm \; 3.4 \; \text{(sys)}] \times 10^{-12}$, consistent with zero violation of the equivalence principle. With a resolution of up to $1.4 \times 10^{-11} \, g$ per shot, we demonstrate a sensitivity to $\eta$ of $5.4 \times 10^{-11}\,/\sqrt{\text{Hz}}$.  
\end{abstract}
\maketitle

The equivalence principle (EP), which posits that all gravitational effects disappear locally \cite{Pauli1921},  is the foundation of general relativity \cite{Weinberg1972} and other geometric theories of gravity. Most theoretical unification attempts that couple gravity to the Standard Model lead to EP violations \cite{Damour1996}. In addition, tests of the equivalence principle search for perturbations of geometric gravity and are sensitive to exotic interactions \cite{Graham2016,Hees2018} that couple differently to the test masses. These tests are complementary to searches for large-scale variations of unknown fields \cite{Arvanitaki2018} and are carried out with local probes that can be precisely controlled. 

EP tests are often characterized by the E\"otv\"os parameter $\eta$, which is the relative acceleration of the test masses divided by the average acceleration between the test masses and the nearby gravitational source.  With classical accelerometers, EP violation has been constrained to $\eta < 1.8 \times 10^{-13}$ by torsion balances in a laboratory setting \cite{Schlamminger2008} and to $\eta < 1.3 \times 10^{-14}$ by the concluded space mission MICROSCOPE \cite{Touboul2017}.  

We perform an equivalence principle test by interferometrically measuring the relative acceleration of freely falling clouds of atoms.  Atom clouds are well-suited test masses because they spend $99.9\%$ of the interrogation time in free fall and the remainder in precisely controlled interactions with the interferometry lasers.  In addition, atoms have uniform and well-characterized physical properties. 
Compared to classical tests, atom-interferometric (AI) EP tests are influenced by different sources of systematic error \cite{Hogan2009}.  AI EP tests can be performed between isotopes that differ only in neutron number, and quantum tests are especially sensitive to particular violation mechanisms \cite{Goklu2008}.  However, previous AI EP tests \cite{Bonnin2013,Schlippert2014,Tarallo2014,Zhou2015} have been limited to $\eta < 3 \times 10^{-8}$ in dual-species comparisons \cite{Zhou2015} and $\eta < 1.4 \times 10^{-9}$ in comparisons between ground states of a single species \cite{Rosi2017b}, largely due to a lack of sensitivity compared to classical experiments. 

In this work, we report an atom-interferometric test of the equivalence principle between $^{85}$Rb and $^{87}$Rb with $\eta = [1.6\; \pm\; 1.8\; \text{(stat)}\; \pm \; 3.4 \; \text{(sys)}] \times 10^{-12}$, consistent with zero violation at the $10^{-12}$ level.  This result improves by four orders of magnitude on the best previous dual-species EP test with atoms \cite{Zhou2015}.  We achieve high sensitivity by utilizing a long interferometer time $T$ and a large momentum splitting between interferometer arms.  With a resolution of $1.4 \times 10^{-11} \, g$ per shot and 15\,s cycle time, the interferometer attains the highest sensitivity to $\eta$ of any laboratory experiment to date \cite{Schlamminger2008}.

The relative acceleration between $^{85}$Rb and $^{87}$Rb is measured with a dual-species atom interferometer.  The experimental apparatus is described in \cite{Overstreet2018}.  We prepare ultra-cold clouds of $^{85}$Rb and $^{87}$Rb by evaporative cooling in a magnetic trap. The subsequent magnetic lensing sequence lowers the horizontal kinetic energies to $25\, \text{nK}$ but introduces a $1.8\, \text{mm}$ vertical offset between the two isotopes. The other kinematic degrees of freedom (DoFs) remain matched. The clouds are then trapped in a vertical 1D optical lattice and accelerated to $13\, \text{m/s}$ in $20\, \text{ms}$. This laser lattice launch transfers an even number of photon recoil momenta $\hbar k$ to the atoms. The final velocity, and therefore the launch height $(\sim 8.5\, \text{m})$, is chosen to match the vertical velocities of the two isotopes. To spatially overlap the clouds, we apply species-selective Raman transitions that kick the two isotopes in opposite directions. After a $77\, \text{ms}$ drift time and removal of untransferred atoms, the Raman transitions are reversed, and the clouds are overlapped to better than $65\, \mu$m. The Raman pulses also provide velocity selection, and the detunings of the Raman pulses allow the average vertical velocity of each isotope to be individually controlled.  

\begin{figure}[h!] 
\centering
\includegraphics[width=\linewidth]{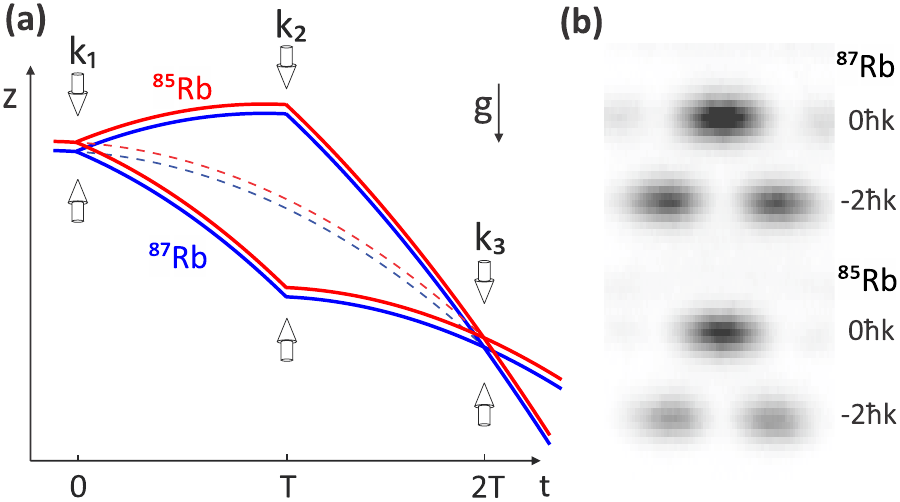} 
\caption{(a) Schematic of simultaneous $^{85}$Rb and $^{87}$Rb interferometer. In pulse zone 1 ($t=0$), each atom cloud is split into two interferometer paths with $\hbar \mathbf{k}_1$ momentum difference. In pulse zone 2 ($t=T$), the paths are reflected toward each other with wavevector $\mathbf{k}_2$. In pulse zone 3 ($t=2T$),  the paths are recombined and interfered with wavevector $\mathbf{k}_3$. The effective wavevectors $\mathbf{k}_1$, $\mathbf{k}_2$, and $\mathbf{k}_3$ differ slightly in orientation and magnitude to create a tailored phase response to kinematic initial conditions.  The midpoint trajectory of each isotope remains essentially unperturbed throughout the interferometer. (b) Single fluorescence image (14.8\,mm x 25.6\,mm) of $^{85}$Rb and $^{87}$Rb output ports ($0\hbar k$ and $-2\hbar k$) with 8$\hbar k$ beamsplitters. The detection fringe allows precise single-shot phase extraction.}
\label{Fig:1}
\end{figure}

The interferometer beamsplitters consist of sequences of two-photon Bragg transitions \cite{Overstreet2018} that transfer $4\hbar k$, $8\hbar k$, or $12 \hbar k$ momentum.  The pulses addressing each interferometer arm are interleaved, and the time interval between successive transitions is $3\,\text{ms}$.  Collectively, these pulses split the clouds symmetrically in the vertical direction.  The symmetric interferometer geometry guarantees that the midpoint trajectory \cite{Antoine2003} of each isotope remains essentially unperturbed.  The interferometer duration $2T$ is $1910\, \text{ms}$, and the maximum wave packet separation for 12$\hbar k$ is $6.9\,\text{cm}$ ($6.7\,\text{cm}$) for $^{85}$Rb ($^{87}$Rb).  After a total drift time of $2.5\, \text{s}$, the output ports (separated by $2\hbar k$ momentum) are imaged with two orthogonal CCD cameras along the horizontal directions.  One isotope is imaged with a time delay of $1\,\text{ms}$ so that the two species can be individually resolved. The phase of each interferometer is given by the population ratio of its output ports.  Fig.~\ref{Fig:1}(a) shows a schematic of the interferometer sequence.

In an EP test configuration, the differential phase between $^{85}$Rb and $^{87}$Rb is close to zero. To distinguish small positive from small negative differential phases, a precise phase offset is needed. By adjusting the angles of the interferometry beams, we imprint a horizontal phase gradient so that each image contains a full interference fringe. This ``detection fringe'' is highly common to both isotopes and allows the contrast and phase of each interferometer to be extracted from a single shot.  Fig.~\ref{Fig:1}(b) shows a fluorescence image in which the detection fringe is visible.

The differential phase shift $\Delta \phi = n\, k\, \Delta g\, T^2$ is proportional to the relative acceleration $\Delta g$ between the atoms. We achieve a single-shot differential phase resolution of $8\, \text{mrad}$ in an $8 \hbar k$ interferometer, which corresponds to a relative acceleration sensitivity of $1.4 \times 10^{-11}\, g$ per shot with duty cycle $15\, \text{s}$.  The observed noise is close to the atomic shot-noise limit with $\sim 10^5$ atoms per interferometer and interference contrast of $70\%$.  In each data run, the initial beamsplitter direction, number of photon recoils $n$ per beamsplitter (4, 8, or 12), detection fringe direction, and imaging order are permuted. The differential phase is averaged over initial beamsplitter direction, detection fringe direction, and imaging order to suppress systematic errors. A full run consists of about $20$ shots in each configuration ($480$ shots total).  The statistical sensitivity is derived from three full runs taken on three separate days.  Throughout the data-taking and analysis process, the EP result was blinded by the addition of an unknown offset to each differential phase measurement.

Systematic errors arise from effects that shift the $^{85}$Rb interferometer phase relative to the $^{87}$Rb phase.  In our experiment, there are three significant sources of systematic error:  differences in kinematic DoFs, differences in the interaction with the electromagnetic field, and imaging errors.  A summary of the systematic errors is presented in Table~\ref{table}.  The most significant systematic effects are described in the text below, and additional errors are discussed in the supplement. 

The relevant kinematic DoFs are the initial position and velocity of each species in the vertical and horizontal directions.  Each species is coupled by its kinematic DoFs to the gravity gradient and to the wavefront of the interferometry beams.  In order to reduce the associated systematic errors, we minimize differences in the kinematic DoFs and suppress the sensitivity of the interferometer to them.  The phase sensitivity to the vertical DoFs can be minimized by adjusting the frequency of the interferometry lasers between each pulse zone, and the phase sensitivity to the horizontal DoFs can be minimized by adjusting the angles of the lasers.

To be concrete, suppose that the average frequency of the interferometry lasers during pulse zone 2 is $f$. Changing the average frequency of pulse zone 1 to $f + \Delta f_1$ adds a phase shift $2 \pi /c \; n\, \Delta f_1\, z$, where $n$ is the number of photon recoils and $z$ is the initial vertical position.  Similarly, if the average frequency of pulse zone 3 is $f + \Delta f_3$, the added phase shift $2 \pi /c\; n\, \Delta f_3\, (z + 2 v_{z} T)$ depends on the initial vertical velocity $v_z$.  Gravity gradients cause systematic errors that are proportional to the vertical initial conditions \cite{Asenbaum2017,Overstreet2018}.  By choosing the appropriate combination of $\Delta f_1$ and $\Delta f_3$, we can simultaneously minimize the phase sensitivity to $z$ and $v_z$, realizing a generalized version of the compensation technique reported in \cite{Roura2017,Overstreet2018}.  To calibrate the pulse zone frequencies, we vertically displace the two isotopes in each DoF and choose $\Delta f_1$ and $\Delta f_3$ to minimize the dependence of the differential phase on the displacement.

An analogous technique can be used to suppress the sensitivity of the interferometer to horizontal DoFs.  Suppose that the beam angle during pulse zone 2 is $\theta_2 = 0$. Setting the angle of pulse zone 1 to $\theta_1$ in the $xz$ plane adds the phase shift $n k\, \theta_1\, x$, where $x$ is the initial horizontal position.  Likewise, the angle $\theta_3$ in pulse zone 3 adds the phase shift $n k\, \theta_3\, (x + 2 v_x T)$, where $v_x$ is the initial horizontal velocity. An appropriate choice of angles in each horizontal direction provides complete compensation of linear phase gradients from the interferometer wavefront.  Such phase gradients arise due to the rotation of the Earth \cite{Sugarbaker2013}.

We control the angle of the interferometry lasers in each pulse zone by setting the angle of the mirror that retroreflects them.  This angle is adjusted during the interferometer to undo the velocity-dependent phase that would otherwise be imprinted by the Earth's rotation.  To calibrate the rotation rate,  we add and subtract an additional velocity-dependent phase in a $^{87}$Rb interferometer and null the fringe frequency difference \cite{Sugarbaker2013}. This procedure suppresses phase shifts proportional to horizontal velocity by a factor of $1000$.  For the EP test, we imprint a horizontal phase that is proportional to the detected position (a combination of initial position and initial velocity; see supplement for additional information). The use of a detection fringe for phase readout avoids systematic errors from initial horizontal displacements between the isotopes.

\begin{figure}[h!]
\centering
\includegraphics[width=\linewidth]{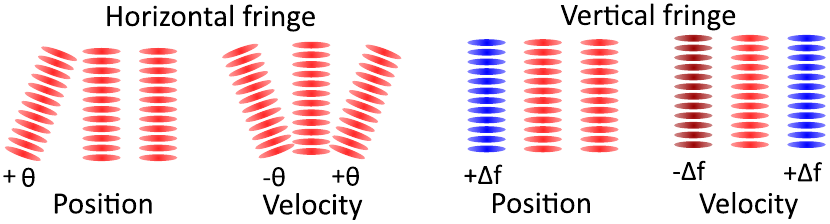}
\caption{Compensation method for kinematic DoFs. Changing the angle (frequency) of pulse zone 1 creates a horizontal (vertical) position-dependent phase. Changing the angle (frequency) of pulse zone 1 by $-\theta$ ($-\Delta f$) and pulse zone 3 by $\theta$ ($\Delta f$) creates a horizontal (vertical) velocity-dependent phase. The angle and frequency steps are calibrated to eliminate phase sensitivity to kinematic DoFs.}
\label{fig:fringes}
\end{figure}

These compensation techniques allow us to suppress all phase shifts that arise from linear horizontal or vertical phase gradients.  However, the atom clouds also have a finite width that can couple to higher-order horizontal wavefront perturbations. To bound this effect, we correlate the cloud width with the differential phase and measure the phase difference between the middle and the edge of each cloud.

To extract the relative acceleration between $^{85}$Rb and $^{87}$Rb, we compare the differential phase of interferometers with beamsplitter momentum $n \hbar k$, where $n \in \{4, 8, 12\}$. The relative acceleration is given by the linear dependence of the differential phase on $n$.  We vary $n$ by adding additional $2\hbar k$ pulses to each pulse zone.  This approach eliminates systematic errors that arise from the finite duration and detuning of the initial and final $\pi/2$ beamsplitter pulses \cite{Antoine2006,Borde1999}, which are the same for all $n$.  
\begin{figure}[h!]
\centering
\includegraphics[width=\linewidth]{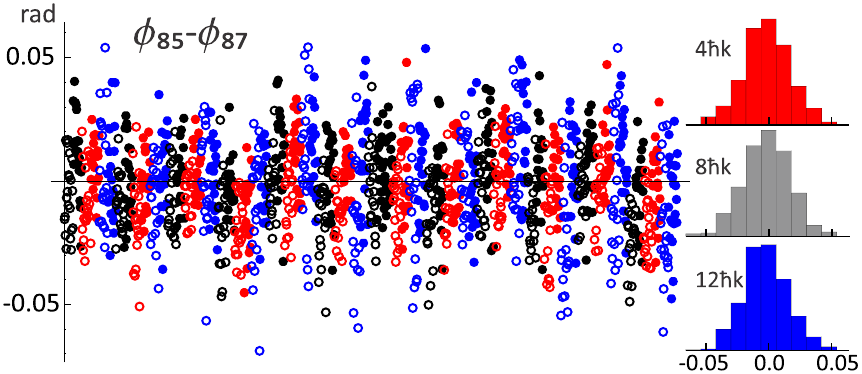}
\caption{EP data (1481 shots).  Left: time series of the interferometer phase difference $\phi_{85}-\phi_{87}$, color-coded by interferometer order ($4\hbar k$ red, $8\hbar k$ gray, $12\hbar k$ blue).  Hollow (solid) points represent measurements with initial beamsplitter direction down (up). Right:  Histogram of the phase difference as a function of interferometer order, averaged over beamsplitter direction, detection fringe direction, and imaging order.}
\label{fig:data}
\end{figure}

The direction of the initial and final beamsplitters, set by the frequency difference between interferometry beams, determines the momenta of the interferometer output ports (either $\{2\hbar k, 0\hbar k\}$ or $\{0\hbar k, -2\hbar k\}$ with respect to the launched clouds). We switch the beamsplitter direction to identify phase shifts that scale with $k^2$.  Such phase shifts arise due to parasitic recoil interferometers \cite{Altin2013} that are caused by imperfect transfer efficiency. The phase of a recoil-sensitive interferometer scales as $\hbar k^2 T/m$, and the dependence on the mass $m$ creates a systematic error in the EP measurement.  When the beamsplitter direction is reversed, the recoil phase shifts change sign relative to the acceleration-induced phase shift, allowing the two effects to be distinguished.

Electromagnetic interactions cause significant systematic errors in two ways.  First, a differential acceleration arises from off-resonant forces (AC-Stark shifts) induced by the interferometry lasers.  This effect is reduced by using an optical spectrum that suppresses the shifts of the $^{87}$Rb $F = 2$ and $^{85}$Rb $F = 3$ states.  To lowest order in the wavepacket separation, the differential Stark shift $\Delta \phi_S$ in our interferometer geometry is given by $\Delta \phi_S = \left(f_{85}/m_{85} -f_{87}/m_{87} \right) 2 \pi (n - 1) \beta \left(n \hbar k T \right)$, where $\beta$ is the fractional intensity gradient at the height of the middle pulse zone and $f_i$ is a factor that characterizes the Stark shift suppression of isotope $i$.  The interferometry beams have a $1/e^2$ radius of $2\,\text{cm}$ and are retroreflected at an angle of $0.5\, \text{mrad}$ to avoid etaloning, which creates a fractional intensity gradient $\beta = 1 \times 10^{-3}/\text{(5 cm)}$.  Reducing the differential Stark shift to $\sim 1\,\text{mrad}$ in a $12\hbar k$ interferometer requires $f_i \sim 10^{-2}$.    

Our optical spectrum is designed to achieve $f_i \sim 10^{-2}$ for both isotopes simultaneously.  The $780\; \text{nm}$ spectrum of each interferometry beam is created by frequency doubling a $1560\, \text{nm}$ laser that is phase modulated at $30 \,\text{GHz}$.  After doubling, the spectrum consists of two strong sidebands and a highly suppressed carrier with relative intensity $\sim 10^{-3}$. The carrier frequency is positioned between the $^{85}$Rb $F=3 \longrightarrow F'$ transitions and the $^{87}$Rb $F=2 \longrightarrow F'$ transitions, which are separated by $1\, \text{GHz}$.  The blue-detuned sidebands are used to drive Bragg transitions, and the red-detuned sidebands compensate the optical forces from the blue-detuned sidebands \cite{Kovachy2015a}.  
To bound the magnitude of the residual AC-Stark effect, we add additional off-resonant pulses to an $8\hbar k$ interferometer.  We observe no statistically significant differential phase shift with these additional pulses, which implies that residual AC-Stark effect induces a differential acceleration below $2.7 \times 10^{-12}\, g$.

Second, the two isotopes are differentially accelerated by magnetic forces.  This effect is reduced by creating a nearly uniform magnetic field in the interferometry region via a solenoid coil and three layers of magnetic shielding \cite{Dickerson2012}. The atoms enter the shielded region in the magnetically sensitive states $\ket{F=3, m_F=3}$ ($^{85}$Rb) and $\ket{F=2, m_F=2}$ ($^{87}$Rb). A series of microwave pulses transfers the atoms to the magnetically insensitive states $\ket{F=3, m_F=0}$ ($^{85}$Rb) and $\ket{F=2, m_F=0}$ ($^{87}$Rb). Nevertheless, the second-order Zeeman effect causes a phase shift $\phi_i = -2(\hbar/m_i) \alpha_i B (\partial_z B) k T^2$ of each isotope, where $\alpha_i$ is the second-order Zeeman coefficient of isotope $i$, $B$ is the magnetic field magnitude, $\partial_z B$ is the vertical magnetic field gradient, and $m_i$ is the mass of isotope $i$. To measure $\partial_z B$, we compare the phase of a $^{87}$Rb interferometer in state $\ket{F=2, m_F=1}$ with the phase of a $^{87}$Rb interferometer in state $\ket{F=2, m_F=0}$. At $B = 41\, \text{mG}$, the $\ket{F=2, m_F=1}$ interferometer has an increased sensitivity to magnetic field gradients by five orders of magnitude. The magnetic field gradient in the interferometry region averages to $(-0.41 \pm 0.036)\, \text{mG/m}$, which corresponds to a differential acceleration of $(5.9 \pm 0.5) \times 10^{-12}\, g$ in the EP measurement.

The phase of each interferometer is encoded in the spatial position of the imaged detection fringe.  Therefore, imaging differences between the isotopes can give rise to systematic errors.  To image both species in a single CCD frame, the fluorescence light for one species is delayed by $1\,\text{ms}$, during which the two isotopes drift apart by $1\, \text{cm}$.  The CCD axis is misaligned with respect to the drift direction by $4\,\text{mrad}$, which causes a differential phase shift of $40\,\text{mrad}$.  To eliminate this phase shift, we rotate the camera images in software.  We also switch the imaging order and reverse the direction in which the detection fringe is imprinted, each of which changes the sign of imaging-related phase shifts relative to the EP signal.  The imaging order and detection fringe direction can each be reversed with fidelity $> 0.99$. Together, these reversals ensure that imaging effects do not contribute significantly to the systematic error.

\begin{table}[h!]
\centering
\begin{tabular}{ l c c }

 Parameter & Shift & Uncertainty \\
\hline
Total kinematic & 1.5 & 2.0 \\
\; \; $\Delta z$ &  & 1.0 \\
\; \; $\Delta v_z$  & 1.5 & 0.7 \\
\; \; $\Delta x$ & & 0.04 \\
\; \; $\Delta v_x$ & & 0.04 \\
\; \; $\Delta y$ & & 0.2 \\
\; \; $\Delta v_y$ & & 0.2 \\
\; \; Width & & 1.6 \\
& & \\
AC-Stark shift & & 2.7 \\
Magnetic gradient & -5.9 & 0.5 \\
Pulse timing & & 0.04 \\
Blackbody radiation & & 0.01 \\
& & \\
\hline
Total systematic & -4.4 & 3.4 \\
Statistical &  & 1.8 \\
\hline
\end{tabular}
\caption{Error budget in units of $10^{-12}\,g$.  The parameter $\Delta z$ ($\Delta v_z$) includes all errors that are linearly proportional to the initial vertical position (velocity) difference between the two isotopes.  Likewise, $\Delta x$ ($\Delta v_x$) includes all errors proportional to the initial position (velocity) difference in the detection fringe direction, and $\Delta y$ ($\Delta v_y$) includes all errors proportional to the initial position (velocity) difference in the orthogonal horizontal direction.  See main text and supplement for descriptions of other systematic errors.  All uncertainties are $1\sigma$.}
\label{table}
\end{table}

We have tested the equivalence principle between $^{85}$Rb and $^{87}$Rb at the level of $10^{-12}\, g$.  The result is consistent with $\eta = 0$, which places generic constraints on new interactions that would differentially accelerate the two isotopes.  The systematic uncertainty is primarily limited by the AC-Stark shift, which can be reduced with an improved laser system that operates at larger single-photon detuning. Such a system would also allow the momentum transfer in each pulse zone to be increased, improving the single-shot sensitivity and reducing the time required to characterize systematic errors.  Uncertainties associated with linear kinematic errors can be reduced to arbitrarily small values by decreasing the statistical uncertainties in the calibration of the compensating frequencies and angles.  The shift due to the magnetic gradient can be reduced by tailoring the current through each segment of the solenoid within the magnetic shield.  By demonstrating that the high sensitivity of large-area atom interferometers can be utilized in precision measurement applications, this work provides a proof of concept for future AI EP tests in space \cite{Aguilera2014} and gravitational wave detectors \cite{Canuel2018,Coleman2018}.

We acknowledge funding from the Defense Threat Reduction Agency, the Office of Naval Research, and the Vannevar Bush Faculty Fellowship program.  We thank Robin Corgier, Salvador Gomez, Jason Hogan, Tim Kovachy, Remy Notermans, and Stefan Seckmeyer for their assistance with this work.

\bibliographystyle{apsrev4-1}
\bibliography{EP-bib}

\end{document}


\title{Atom-interferometric test of the equivalence principle at the $10^{-12}$ level \\ Supplemental Material}

\author{Peter Asenbaum}
\thanks{These authors contributed equally to this work.}
\author{Chris Overstreet}
\thanks{These authors contributed equally to this work.}
\author{Minjeong Kim}
\author{Joseph Curti}
\author{Mark A. Kasevich}
\affiliation{Department of Physics, Stanford University, Stanford, California 94305}
\date{June 25, 2020}
\maketitle

\section{Data analysis}

\begin{figure}[h!]
\centering
\includegraphics[width=\columnwidth]{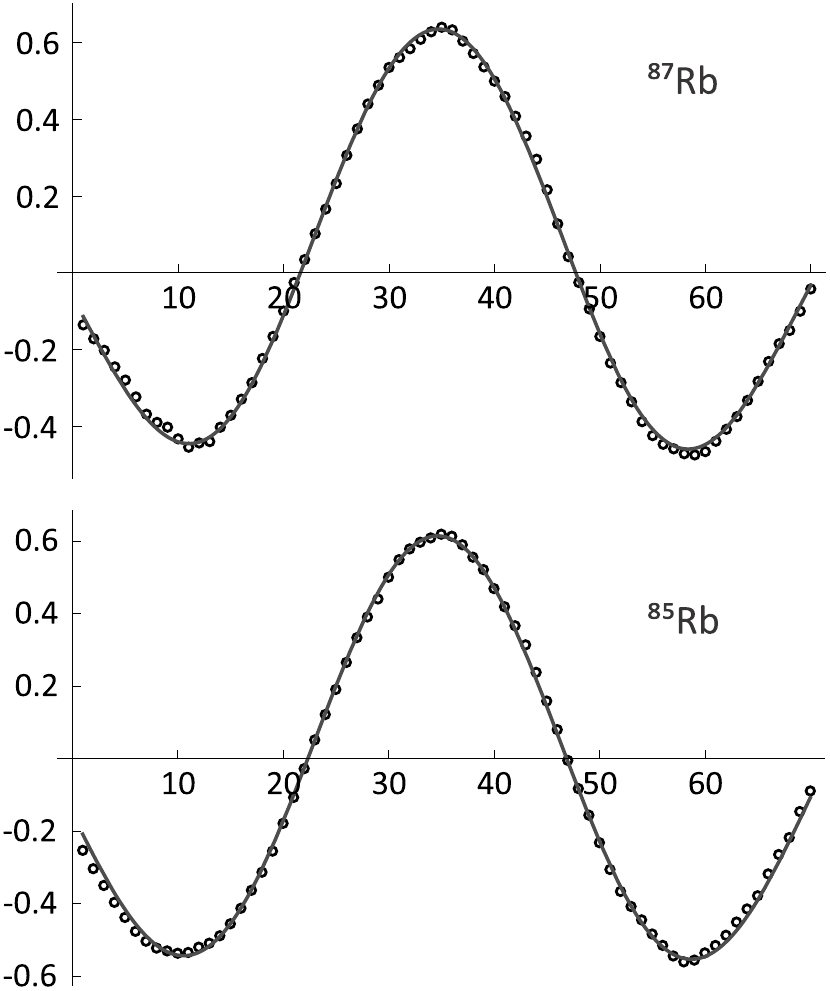}
\caption{Asymmetries and fits of a single shot.  The $x$ axis represents the horizontal pixel number on the CCD array (1 pixel = $170\, \mu\text{m}$).}
\label{fig:asym}
\end{figure}

Each dual-species interferometer image contains four output ports (two for each isotope) separated vertically by $\sim 2.5\,\text{mm}$ on the CCD array.  To analyze the images, we bin each port in the vertical direction and construct the asymmetry 
\begin{equation}
    A(x) = \frac{P_1(x) - P_2(x)}{P_1(x) + P_2(x)}
\end{equation}
of each interferometer, where $P_i(x)$ is the number of counts in the $i^\text{th}$ port as a function of horizontal pixel number.  Ideally, the asymmetry is given by 
\begin{equation} \label{Eq:simpleInt}
    A_1(x) = c \cos \left[k_x (x - x_0) + \phi\right]
\end{equation}
where $c$ is the interferometer contrast, $k_x$ is the detection fringe frequency, $x_0$ is a reference point chosen to be in the center of the port, and $\phi$ is the interferometer phase.  Our asymmetry model incorporates two additional effects.  First, the interferometer contrast varies slowly as a function of horizontal position due to Rabi frequency inhomogeneity.  Second, the photon collection efficiency of the imaging system varies as a function of vertical position.  We therefore use the model
\begin{equation}
    A_2(x) = \frac{(1 - r) + (1 + r) \exp \left[-(x - x_0)^2 / (2 \sigma^2) \right] A_1(x)}{(1 + r) + (1 - r) \exp \left[-(x - x_0)^2 / (2 \sigma^2) \right] A_1(x)}
\end{equation}

where $\sigma$ characterizes the length scale of the contrast variation and $r$ is the collection efficiency ratio between the two output ports.  As $r \longrightarrow 1$ and $\sigma/(x - x_0) \longrightarrow \infty$, this expression reduces to Eq.~\ref{Eq:simpleInt}.  A typical fit of this model to the data is shown in Fig.~\ref{fig:asym}.  We verified that the fitted interferometer phase $\phi$ is insensitive to the precise form of the model.  

Each interferometer is fit independently, and the relevant information from each image is the differential phase $\phi_\text{85} - \phi_\text{87}$ between the two interferometers.  The EP signal is the linear dependence of the differential phase on the number of photon recoils $n$, where $n \in \{4, 8, 12\}$.  This signal was blinded throughout the data-taking and analysis process by adding an unknown offset $n/4 \cdot \phi_\text{blind}$ to each differential phase, where $\phi_\text{blind}$ was randomly chosen from a Gaussian distribution with mean zero and standard deviation $2 \, \text{mrad}$ (corresponding to a $1\sigma$ fractional acceleration difference of $7 \times 10^{-12}$).  After removing the blind, the result was published without taking additional data or altering the data analysis in any way.      

\section{Detection fringe}

The angles of the interferometry beams are adjusted to create a horizontal phase shift that is proportional to the position at the time of detection.  To fix notation, suppose that the initial beamsplitter, mirror, and final beamsplitter pulses occur at times $t = 0$, $t = T$, and $t = 2T$, respectively, and that there is a time interval $\tau$ between the final beamsplitter pulse and the fluorescence detection.  Then the horizontal phase shift $\phi_h$ is given by 
\begin{equation}
\phi_h = n k \theta_1 x + n k \theta_3 (x + v_x \cdot 2T).
\end{equation}
Here $x$ is the initial horizontal position, $v_x$ is the initial horizontal velocity, $\theta_1$ and $\theta_3$ are the angles of the initial and final beamsplitter pulse, respectively, and we have chosen coordinates so that the mirror pulse angle is zero.  To create a detection fringe, we choose $\theta_1$ and $\theta_3$ so that 
\begin{equation}
\phi_h = \alpha k\, \big(x + v_x \cdot (2T + \tau)\big) 
\end{equation}
where $\alpha$ is a dimensionless number that sets the detection fringe frequency.  Comparing the coefficients of $x$ and $v_x$, we have 
\begin{equation}
\theta_1 = -\frac{\alpha}{n}\cdot \frac{\tau}{2T}
\end{equation}
and
\begin{equation}
\theta_3 = \frac{\alpha}{n}\cdot \frac{2T + \tau}{2T}.
\end{equation}
We use this technique to implement a detection fringe in large momentum transfer interferometers, taking into account the time interval between sequential pulses in each pulse zone.  

\section{Systematic errors} 

\subsection{I. Kinematic effects}

We suppress the linear phase gradients associated with all six kinematic DoFs by adjusting the frequency and angles of each pulse zone, as described in the main text.  To calibrate the compensation for a particular DoF, the two clouds are differentially displaced in that DoF, and the resulting differental phase shift is observed.  The initial position difference in the vertical direction is adjusted by varying the time between the Raman velocity selection pulses, during which the two isotopes have a differential velocity of $2\hbar k/m$.  The initial velocity difference in the vertical direction is adjusted by varying the frequencies of the Raman velocity selection pulses.  To change the initial position and velocity differences in the horizontal directions, we adjust the position of the optical lattice that is used to launch the atoms.  During the EP measurement, the uncertainties in the differential kinematic DoFs are $(65, 20, 20)\, \mu\text{m}$ for $(\Delta z, \Delta x, \Delta y)$ and $(60, 50, 40)\; \mu\text{m/s}$ for $(\Delta v_z, \Delta v_x, \Delta v_y)$.

There is a nonzero shift associated with the vertical velocity difference because the frequency of the Raman velocity selection pulse for each species was chosen to optimize the transfer efficiency of the first beamsplitter pulse, which selects distinct velocity classes of the two isotopes due to the recoil velocity difference.  This velocity difference multiplies with a residual velocity-dependent phase gradient that was observed after the EP data-taking was completed.    

In order to suppress phase shifts proportional to horizontal velocity by a factor of 1000, the beam angles are set with a precision of $50\,\text{nrad}$.  The angle of the retroreflection mirror is controlled by a feedback loop.  A position-sensitive detector measures the position of a laser that is reflected off of the back of the mirror, and the mirror angle is adjusted by changing the voltage across the three piezoelectric actuators on which the mirror is mounted.   

Since the two species are fit with the same weighting function in the $x$ direction, the differential phase is insensitive to quadratic wavefront perturbations on that axis.  However, a quadratic wavefront perturbation in the $y$ direction induces a differential phase due to the cloud width difference.  We bound the magnitude of this effect by correlating the width difference in the $y$ direction with the differential phase.  In addition, we estimate the phase shift in the $y$ direction from the observed quadratic dependence of the phase in the $x$ direction.  This estimate provides the uncertainty quoted in Table I of the main text.      

\subsection{II. AC-Stark shift}

\begin{figure}[h!]
\centering
\includegraphics[width=\columnwidth]{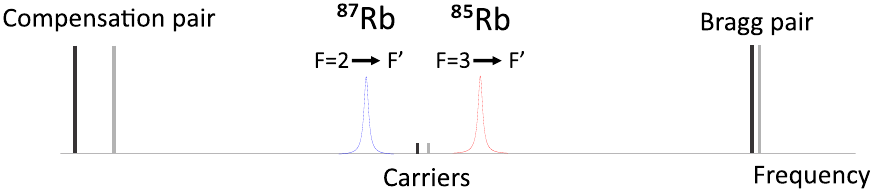}
\caption{Optical spectrum.  The carriers, located between the $^{85}$Rb and $^{87}$Rb transitions, are separated by $158\,\text{MHz}$.  The frequency difference between the blue sidebands (Bragg pair) is adjusted to be resonant with the two-photon Bragg transition for each pulse.  The red sidebands (Compensation pair) are separated by an additional $316\, \text{MHz}$ and do not drive Bragg transitions.  AC-Stark compensation for both isotopes is achieved by controlling the intensity ratio between carriers and sidebands.  Figure is not to scale.}
\label{fig:universe}
\end{figure}

In a Bragg interferometer, each atom remains in a single internal state throughout the interferometry sequence, so the interferometer is not sensitive to relative energy shifts between internal states (as in \cite{Zhou2015}).  However, the AC-Stark effect induces a differential phase shift if the laser intensity on the interferometer arms differs during the middle pulse zone.  As described in the main text, the optical spectrum of the interferometry lasers is designed to reduce the magnitude of the AC-Stark shift.    
Fig.~\ref{fig:universe} contains a schematic of the optical spectrum.  About $25\%$ of the power of each beam is transferred into the second-order sidebands.

We compute the Stark shift of the measured spectrum and control the carrier frequencies, carrier amplitudes, and sideband asymmetries to maintain Stark shift compensation.  The differential Stark shift depends primarily on the amplitudes of the carriers, which are ideally suppressed to a fractional intensity of $7.5 \times 10^{-4}$ compared to the sidebands.  

To bound the systematic error associated with the AC-Stark effect, we operate a dual-species $8\hbar k$ interferometer with twice as many pulses in each pulse zone.  The additional pulses (``Stark shift zones'') are detuned from the two-photon resonance and do not cause Bragg transitions, but they double the magnitude of the residual Stark shift.  The difference in the differential phase of the $8\hbar k$ interferometer with and without Stark shift zones is below $0.96\, \text{mrad}$, which limits the systematic uncertainty in the EP measurement to $2.7 \times 10^{-12}$.  

\subsection{III. Magnetic gradient}

As described in the main text, a magnetic field gradient induces a differential acceleration between the two isotopes through the second-order Zeeman effect.  We characterize the magnetic gradient with a $4\hbar k$ $^{87}$Rb gradiometer in which one interferometer is in the magnetically sensitive state $\ket{F=2, m_F=1}$.  The other interferometer is in the magnetically insensitive state $\ket{F=2, m_F=0}$ and acts as a phase reference.  The two interferometers are spatially overlapped to within $1\, \text{cm}$.   

\begin{figure}[h!]
\centering
\includegraphics[width=\columnwidth]{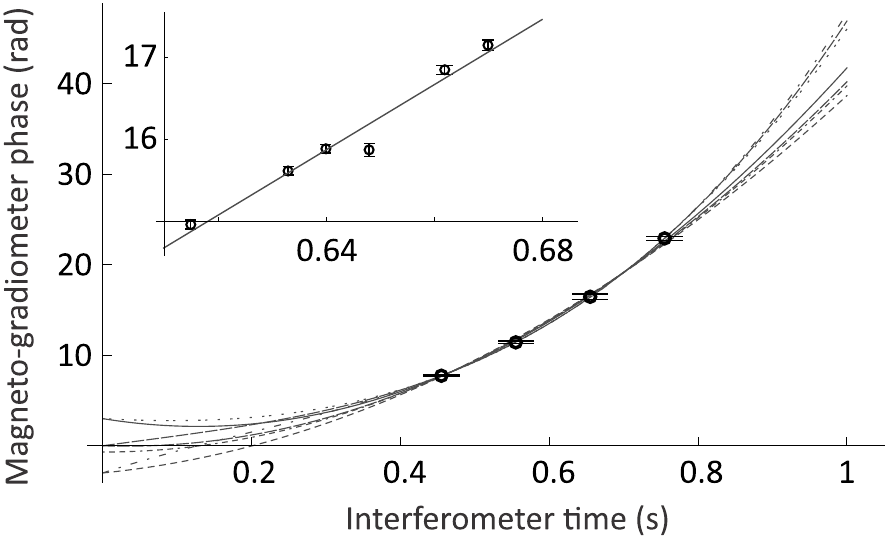}
\caption{Magnetically sensitive interferometry.  Main plot: phase difference between magnetically sensitive and insensitive interferometer as a function of interferometer time.  The extrapolations used to estimate the phase shift at $T = 955\, \text{ms}$ are also shown.  Inset: measurements with small timing offsets near $T = 655\, \text{ms}$.}
\label{fig:magsan}
\end{figure}

Fig.~\ref{fig:magsan} shows the differential phase of the $^{87}$Rb gradiometer as a function of interferometer time $T$.  The gradiometer phase varies by much more than $2 \pi$ as the interferometer time is increased.  To determine the scale of the phase variation, we took data with small timing offsets near $T = 655\, \text{ms}$.  

The sub-quadratic scaling of the gradiometer phase with $T$ indicates that the magnetic gradient varies over the interferometry region, and the gradiometer phase is proportional to an average of the magnetic gradient on the interferometer trajectory (see supplement of \cite{Overstreet2018} for an analogous calculation).  To estimate the average magnetic gradient for the EP measurement, we fit several low-order polynomials to the measured gradiometer phase and extrapolate to $T = 955\, \text{ms}$.  Specifically, we use the model $y = a x^3 + b x^2 + c x + d$, with various combinations of $\{a, b, c, d\}$ as free parameters.  The quoted uncertainty is derived from the range of values predicted by these extrapolations.          

\subsection{IV. Pulse timing}

The phase of an atom interferometer generally depends on the time between each interferometer pulse.  We operate the interferometer with a $\sim 3\,\text{ms}$ delay between pulses, which improves contrast and suppresses spurious interferometers associated with imperfect transfer efficiency.  The pulses have a Gaussian temporal profile and a maximum Rabi frequency of $20\,\text{kHz}$.  The pulse area is controlled with a precision of $0.15\%$ by a photodiode that monitors the intensity of a sampled beam. A timing error $\delta T$ in a single pulse produces a differential phase shift of $\sim 2k(\Delta v_z + \Delta r)\delta T$, where $\Delta r =  \hbar k/m_{85} - \hbar k/m_{87}$ is the recoil velocity difference between isotopes.  The term proportional to $\Delta v_z$ is already included in the kinematic error, but the term proportional to $\Delta r$ is an additional source of uncertainty if the timing error is not constant when the beamsplitter direction is reversed.

We bound pulse timing errors by using an oscilloscope to record the intensity of the interferometry lasers as a function of time during the interferometry sequence.  The center of each pulse is identified by fitting the temporal profile of the pulse to a Gaussian.  There is a $40\, \text{ns}$ offset between the expected and observed location of each $\pi/2$ pulse, with the $\pi/2$ pulses occurring earlier than expected.  This offset may be caused by the finite bandwidth of the acousto-optic modulator that is used to control the pulse intensity. The associated phase shift reverses with beamsplitter direction and does not scale with interferometer order, so it does not contribute to the EP uncertainty.  The remaining pulses are correctly timed to within $5\, \text{ns}$, which corresponds to an additional uncertainty of $4 \times 10^{-14}$ in the EP result. 

\subsection{V. Blackbody radiation}

A thermal gradient in the interferometry region creates a blackbody radiation gradient that exerts forces on $^{85}$Rb and $^{87}$Rb \cite{Sonnleitner2013,Haslinger2018}.  Although the two isotopes have the same ground-state polarizability, the resulting differential acceleration is nonzero because of the mass difference.  We monitor the temperature in the interferometry region with a series of thermocouples placed at $\sim 1\, \text{m}$ intervals along the length of the $10\, \text{m}$ vacuum chamber.  The thermal gradient in the vertical direction is below $0.1 \degree \text{C/m}$, and a numerical simulation indicates that the resulting differential acceleration is below $10^{-14}\, g$.  

\subsection{VI. Mean field interaction}

Cold intra- and inter-species collisions of $^{85}$Rb and $^{87}$Rb in the freely falling atom clouds lead to a density-dependent energy shift \cite{Aguilera2014,Debs2011}. If the initial beamsplitter transfer is unbalanced, the two interferometer arms will experience different mean field energies.  The Bragg lasers are $3\%$ less detuned from the $^{85}$Rb resonance than from the $^{87}$Rb resonance, and the resulting Rabi frequency difference creates unbalanced $^{85}$Rb densities along the interferometer arms. At the first pulse zone,  the atom clouds have a horizontal width of $1.5\,\text{mm}$ and a vertical width of $0.7\,\text{mm}$ (FWHM). The associated phase shift is estimated to be $0.2\,\text{mrad}$, nearly independent of interferometer order.  As a function of interferometer order, the density decreases due to spontaneous emission by 1.3\% per $4\hbar k$, which leads to a relative error of less than $8 \times 10^{-15}$ in the EP result.

%